\documentclass[prb,twocolumn,10pt,floatfix]{revtex4}

\usepackage{graphicx}

\begin{document}

\title{Ultrafast Relaxation Dynamics of Hot Optical Phonons in Graphene}

\author{Haining Wang, Jared H. Strait, Paul A. George, Shriram Shivaraman, Virgil B. Shields, Mvs Chandrashekhar, Jeonghyun Hwang, Farhan Rana, Michael G. Spencer}
\affiliation{School of Electrical and Computer Engineering, Cornell University, Ithaca, NY, 14853}
\author{Carlos S. Ruiz-Vargas, Jiwoong Park}
\affiliation{Department of Chemistry and Chemical Biology, Cornell University, Ithaca, NY, 14853}

\begin{abstract}
Using ultrafast optical pump-probe spectroscopy, we study the relaxation dynamics of hot optical phonons in few-layer and multi-layer graphene films grown by epitaxy on silicon carbide substrates and by chemical vapor deposition on nickel substrates. In the first few hundred femtoseconds after photoexcitation, the hot carriers lose most of their energy to the generation of hot optical phonons which then present the main bottleneck to subsequent carrier cooling.  Optical phonon cooling on short time scales is found to be independent of the graphene growth technique, the number of layers, and the type of the substrate. We find average phonon lifetimes in the 2.5-2.55 ps range. We model the relaxation dynamics of the coupled carrier-phonon system with rate equations and find a good agreement between the experimental data and the theory. The extracted optical phonon lifetimes agree very well with the theory based on anharmonic phonon interactions.
\end{abstract}

\maketitle

The performance of most demonstrated and proposed graphene-based electronic and optical devices depends critically on carrier and optical phonon scattering as well as relaxation dynamics \cite{Shepard08,Jena09,Rana08,Avouris09,Mac09}. Optical phonons in graphene can decay in two main ways: i) optical phonons can get absorbed by electrons or holes, or ii) optical phonons can decay into other phonons via anharmonic interactions and lattice defects. In this paper, phonon lifetime refers to the lifetime due to the latter processes, the most important of which is the decay of an optical phonon into two acoustic phonons \cite{Bonini07}. Optical phonon lifetimes in graphene can be estimated from frequency domain measurements, such as Raman spectroscopy. However, the measured Raman linewidths also contain contributions from pure dephasing processes and inhomogeneous broadening \cite{Heinz08}. Direct measurement of the optical phonon lifetime in carbon nanotubes via time-resolved Raman anti-Stokes spectroscopy was performed by Song et.\ al. \cite{Heinz08} and a value of  $\sim$1.1 ps was reported. Theoretical calculations of the hot optical phonon lifetimes in graphene due to anharmonic decay into acoustic phonons were carried out by Bonini et.\ al. \cite{Bonini07}, and values between 2-3 ps were reported for both zone-center ($E_{2g}$) and zone-edge ($A'_{1}$) modes for phonon temperatures in the 500-900 K range. A much longer lifetime value of $\sim$7 ps was measured by Kampfrath et.\ al. \cite{Kampfrath05} via time-resolved measurements of the electronic temperature using optical-pump terahertz-probe spectroscopy. Degenerate optical pump-probe studies of carrier dynamics in epitaxial graphene have been reported previously by the authors and others \cite{Jahan08,Driel09}. In this paper, we use optical pump-probe spectroscopy to study optical phonon lifetimes in graphene, and we also present a theoretical framework for interpreting the results. 
   
Graphene samples used in this work were grown on the silicon and carbon faces of semi-insulating silicon carbide (6H-SiC) wafers by thermal decomposition (epitaxial growth \cite{Heer06}) and also by chemical vapor deposition (CVD) on nickel \cite{Hong09,Reina09}. Samples A and B were grown on the carbon face of SiC and  Sample C was grown on the silicon face of SiC. Sample D was grown by CVD on nickel and then transferred onto a quartz substrate. Sample D was found to have patches in which the number of layers varied between 1 and 3. All samples were characterized via Raman spectroscopy (pump wavelength 488 nm) and optical/IR transmission spectroscopy. The characteristics are recorded in Table I. For spectroscopy, the pump and the probe beams were obtained from a Ti:Sapphire mode-locked laser with a 81 MHz pulse repetition rate and a 780 nm center wavelength. The pulse width (FWHM) at the sample was measured to be $\sim$100 fs. The polarization of the probe was rotated $90\,^{\circ}$ with respect to the pump such that scattered pump light could be filtered out of the probe beam before photodetection using a polarizer and a spatial filter.  The differential transmission (DT) of the probe due to the pump was obtained by chopping the pump and probe at different frequencies and recording the photodetector current at the sum frequency with a lock-in amplifier.

The measured time-resolved DT signals (normalized to their peak values) for all the four samples are shown in Fig.1 (solid circles). In each case, the pump energy was $\sim$14 nJ (fluence $\sim$15 $\mu$J/cm$^{2}$). The photoexcited carrier density was estimated to be $\sim$5$\times$10$^{11}$ cm$^{-2}$ per layer for Samples A through C (SiC substrate), and $\sim$8$\times$10$^{11}$ cm$^{-2}$ per layer for Sample D (quartz substrate). The autocorrelation of the $\sim$100 fs pulses used in the experiments are also shown in Fig.1 (dashed lines) for reference. The measured transients exhibit two distinct time scales. The DT signal decreases rapidly to almost 10$\%$ of its peak value in the first $\sim$400-500 fs, and then it decreases much more slowly afterwards. These results can be qualitatively explained in the following way. We assume that the probe transmission is affected only by the change in the interband absorption caused by the pump pulse. Since the probe transmission is sensitive to the carrier occupation in the conduction and valence bands at energies $\sim$$\hbar \omega/2$ above the Dirac point\cite{Jahan08} (half of the pump photon energy), it is also sensitive to the carrier temperature. Immediately after photoexcitation, the photogenerated electrons and holes thermalize with each other and the existing carriers, thereby acquiring Fermi-Dirac distributions with high temperatures. Recent studies have shown that the thermalization times for the photoexcited carriers in graphite are extremely short and in the 20-40 fs range \cite{Elsaesser09}. Due to the fast electron-hole scattering times, it is also reasonable to assume that the electron and hole temperatures are approximately the same. The thermalized electron and hole distributions then cool down via interaction with the optical phonons. The hot photoexcited carriers lose most of their energy to the optical phonons in the first 500 fs after photoexcitation. This results in the generation of hot optical phonons, which then present a bottleneck to the subsequent cooling of the carriers. The relaxation dynamics of the carriers and the optical phonons are strongly coupled, and experimental data can be interpreted correctly only if this coupling is taken into account. Below, we present a model that considers this coupling, and we use this model to extract the optical phonon lifetime from the data.    

\begin{figure}[tb]
  \centering
    \includegraphics[width=3.2in]{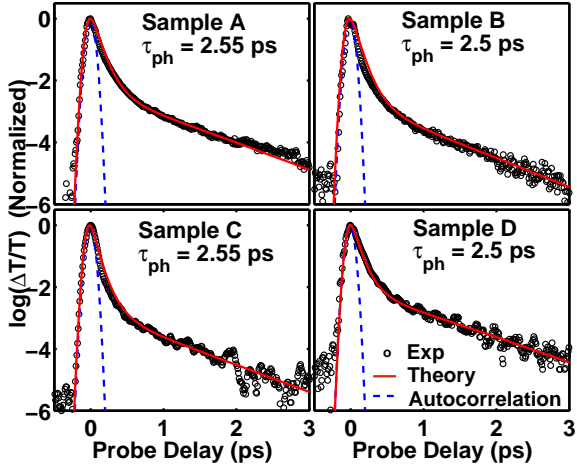}
  \caption{Measured differential probe transmissions (normalized to peak value)(circles). The autocorrelation of the $\sim$100 fs pump and probe pulses (dashed lines), and the best-fit theoretical curves (solid lines) are plotted for different graphene samples. The extracted values of the average optical phonon lifetime $\tau_{ph}$ are indicated for each graphene sample.}  
  \label{Fig1}
\end{figure}

\begin{table}[htb]
\caption{Experimentally determined quantities: Ratio of Raman G-peak and D-peak intensities $I_{G}/I_{D}$, the number of graphene layers $N$, the extracted values of the optical phonon lifetime $\tau_{ph}$ and the average carrier density per layer $n_{o}$.}
\label{summary}
\begin{tabular}{c c c c c}
\hline
Sample &$I_{G}/I_{D}$ &$N$ & $\tau_{ph}$ & $n_{o}$ \\
\hline \hline
A & 20.5 & 55 & 2.55$\pm$0.08 ps & 1$\times$$10^{11}$ cm$^{-2}$ \\
\hline
B &57 &16 & 2.5$\pm$0.08 ps & 1$\times$$10^{11}$ cm$^{-2}$  \\
\hline
C & 9.6 & 2 & 2.55$\pm$0.1 ps &  6$\times$$10^{11}$ cm$^{-2}$ \\
\hline
D & 19 & $\sim$2 & 2.5$\pm$0.12 ps &  7$\times$$10^{11}$ cm$^{-2}$ \\
\hline
\end{tabular}
\end{table}

\begin{figure}[tb]
\centering
\includegraphics[width=3.5in]{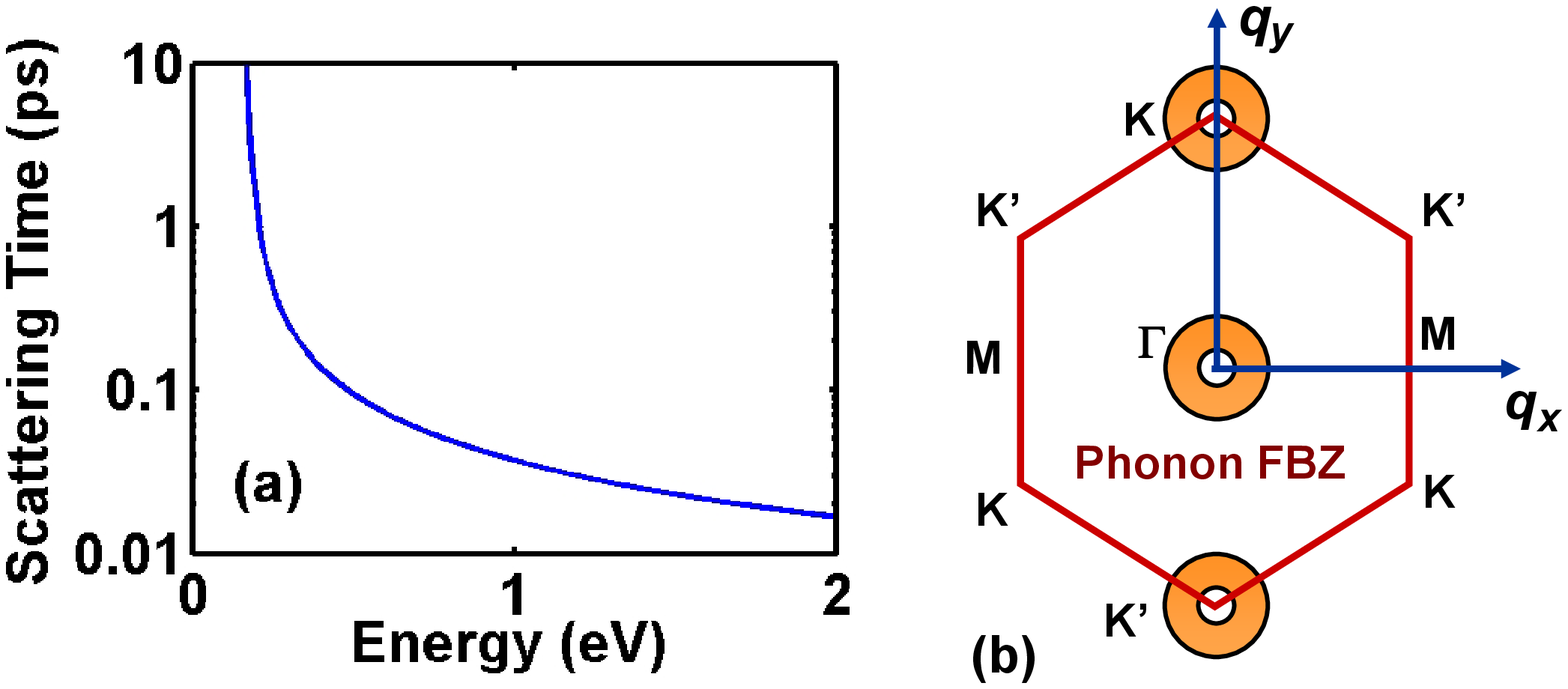}
\caption {(a) The electron intraband scattering time due to optical phonon scattering is plotted as a function of the electron energy \cite{Rana09}. (b) The phonon Brillouin zone is shown together with the annular regions (not to scale) at the $\Gamma$ and K(K') points, in which the optical phonons participate in electron-phonon scattering.}
\label{Fig2}
\end{figure}

\begin{figure}[tb]
  \centering
    \includegraphics[width=3.7in]{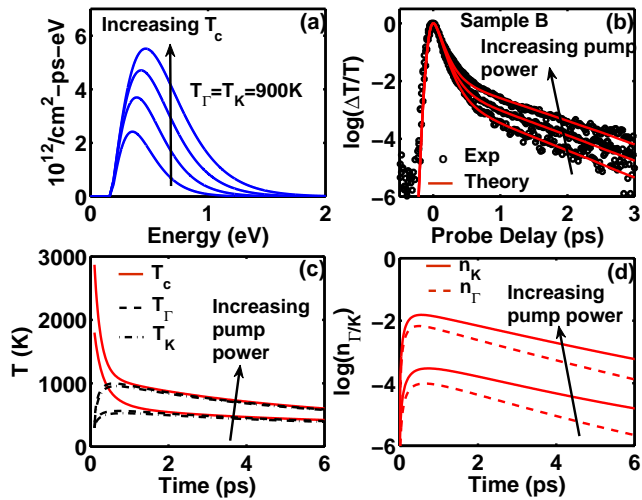}
  \caption{(a) Calculated electron-phonon scattering rate versus electron energy is plotted for values of $T_{c}$ equal to 1250K, 1500K, 1750K, and 2000K. Total electron density is assumed to be $7\times 10^{11}$ cm$^{-2}$. Phonon temperature is fixed at 900K. (b) Measured differential probe transmissions (normalized to the peak value) (circles) and the best-fit theoretical curves (lines) are plotted for different pump pulse energies (2.5, 6, 9.8 nJ). (c) and (d) The calculated electron and phonon temperatures and the phonon occupation numbers are plotted as a function of time after photoexcitation. The photoexcited electron-hole densities are assumed to be  $2\times 10^{11}$ cm$^{-2}$ and  $7\times 10^{11}$ cm$^{-2}$. The equilibrium electron density is assumed to be  $10^{11}$ cm$^{-2}$.}
  \label{Fig3}
\end{figure}

Intraband electron-phonon scattering can be intravalley (due the $\Gamma$-point $E_{2g}$ optical phonons) or intervalley (due to the K(K')-point $A'_{1}$ phonons). Using the matrix elements for electron-phonon scattering given by Rana et. al. \cite{Rana09}, the net optical phonon emission rates,  $R_{\rm \Gamma e}$ and $R_{\rm K e}$ (units:  cm${^{-2}}$s$^{-1}$), due to intraband intravalley and intervalley electron-phonon scattering, respectively, can be expressed as,
\begin{eqnarray}
R_{\rm \Gamma e} & \approx &  9 \frac{       (dt/db)^2         }{\displaystyle \pi \, \rho \, \omega_{\rm \Gamma} \, \hbar^{4} \, v^{4}} \, \int^{\infty}_{\hbar \omega_{\rm \Gamma}}  dE \, E \, \left( E - \hbar \omega_{\rm \Gamma} \right) \nonumber \\
& & \times \left\{ f_{c}(E) \left[1-f_{c}(E-\hbar\omega_{\rm \Gamma})\right]\,\left( 1 + n_{\rm \Gamma} \right) \right. \nonumber \\
& & -  \left.     f_{c}(E-\hbar\omega_{\rm \Gamma}) \left[1-f_{c}(E)\right]\, n_{\rm \Gamma}   \right\} \label{eqr1}
\end{eqnarray} 
The integrand in the above Equation is the distribution in energy of the electron-phonon scattering events. The expression for $R_{\rm K e}$ is the same as above except for the replacement $\rm \Gamma \rightarrow K$. Here, $dt/db$ equals $\sim$45 eV/nm~\cite{Rana09,Piscanec07}, $\rho$ is the density of graphene and equals $\sim 7.6 \times 10^{-7}$ kg/m$^{2}$, $\omega_{\rm \Gamma}$ ($n_{\rm \Gamma}$) and $\omega_{\rm K}$ ($n_{\rm K}$) are the frequencies (average occupation numbers) of the $E_{2g}$ and $A'_{1}$ phonons, respectively, $v=10^{6}$ m/s, and $f_{c}(E)$ is the Fermi-Dirac distribution function for conduction band electrons. Similar expressions as above can be written for the optical phonon emission rates, $R_{\rm \Gamma h}$ and $R_{\rm K h}$, due to hole-phonon intraband scattering. The carrier temperature $T_{c}$ and the optical phonon occupation numbers, $n_{\rm \Gamma}$ and $n_{\rm K}$, obey the coupled rate equations,
\begin{eqnarray}
\frac{d\,T_{c}}{dt} & = & -\frac{ (R_{\rm \Gamma e} + R_{\rm \Gamma h}) \hbar \omega_{\rm \Gamma} + (R_{\rm K e} + R_{\rm K h}) \hbar \omega_{\rm K}}{C_{e} + C_{h}} \label{eq1} \\
\frac{d\,n_{\rm \Gamma}}{dt} & = & \frac{R_{\rm \Gamma e} + R_{\rm \Gamma h}}{M_{\rm \Gamma}} - \frac{n_{\rm \Gamma} - n^{o}_{\rm \Gamma}}{\tau_{ph}} \label{eq2} \\
\frac{d\,n_{\rm K}}{dt} & = & \frac{R_{\rm K e} + R_{\rm K h}}{M_{\rm K}} - \frac{n_{\rm K} - n^{o}_{\rm K}}{\tau_{ph}} \label{eq3}
\end{eqnarray}
Here, $C_{e}$ and $C_{h}$ are the electron and hole heat capacities and $\tau_{ph}$ is the average lifetime of the $E_{2g}$ and $A'_{1}$ optical phonons. $n^{o}_{\rm \Gamma}$ and $n^{o}_{\rm K}$ are the phonon numbers in equilibrium (at T=300K). $M_{\rm \Gamma}$ and $M_{\rm K}$ are the number of phonon modes (per unit area) at the $\Gamma$ and K(K') points, respectively, that participate in carrier-phonon interaction. The values of $M_{\rm \Gamma}$ and $M_{\rm K}$ can be estimated as follows. At the $\Gamma$-point, only phonons with wavevector magnitudes $q$ in a certain range (dictated by momentum and energy conservation) participate in carrier-phonon interaction. If an electron with wavevector $k$ emits an optical phonon, then the smallest value of $q$ is $\omega_{\rm \Gamma}/v$ and most of the emitted phonons have wavevectors smaller than $k$. For carrier densities in the $10^{11}-10^{12}$ cm$^{-2}$ range, and carrier temperatures in the 300-3000K range, most of the electrons that emit optical phonons have energies less than $E_{max} \approx 0.8$ eV (see Fig.3(a)). Therefore, the largest value of $q$ can be approximated by $E_{max}/\hbar v$. These estimates have been verified by detailed calculations using wavevector dependent distributions for the phonons \cite{Butscher07}. The value of $M_{\rm \Gamma}$ then corresponds to the number of phonon modes that are in the annular region (see Fig.2(b))defined by the minimum and maximum values of $q$ and equals $2\times[(E_{max}/\hbar v)^{2} - (\omega_{\rm \Gamma}/v)^{2}]/4\pi$. The factor of 2 in the front stands for the two degenerate LO and TO phonon bands at the $\Gamma$-point. Similarly, $M_{\rm K}$ equals $2\times[(E_{max}/\hbar v)^{2} - (\omega_{\rm K}/v)^{2}]/4\pi$. Here, the factor of 2 in the front stands for the two degenerate phonon valleys at the K and K' points. Since the optical phonon emission time for an electron at an energy $\hbar \omega/2\approx 0.8$ eV above the Dirac point is less than $\sim$50 fs (see Fig.2(a)), the initial carrier temperature is computed by assuming that during the first 100 fs after photoexcitation all the photogenerated carriers emit $\sim$2 optical phonons. The set of coupled Equations (\ref{eq1})-(\ref{eq3}) were solved to obtain the carrier temperature and the phonon numbers as a function of time. It is assumed that no significant carrier recombination occurs in the first few picoseconds \cite{Paul08}. The carrier distributions were then used to compute the optical transmission using the expressions given by Dawlaty et.\ al. \cite{Jahan08} and the results were convolved with a probe pulse assumed to be Gaussian with a FWHM of 100 fs. Theoretical results obtained are plotted in Fig.1 (solid lines). The agreement with experimental data is seen to be very good, and both the fast and slow time scales in the DT signal are correctly reproduced. It needs to be pointed out here that the only two fitting parameters in the model are the phonon lifetime, $\tau_{ph}$, and the equilibrium carrier density per layer, $n_{o}$. Their extracted values are given in Table I. The error bars in Table I for the extracted values of $\tau_{ph}$ correspond to the spread in the data obtained when the measurements were repeated many times. The extracted values of $\tau_{ph}$ are not sensitive to the extracted values of $n_{o}$ nor to the photoexcited electron-hole densities. Fig.3(b) shows the agreement between the theory and the data for pump pulse energies of 2.5, 6, and 9.8 nJ (fluences 6.2, 15, and 24.5 $\mu$J/cm$^{2}$). In the model, the photoexcited carrier density was scaled with the pump pulse energy. In each case, the average best fit value for $\tau_{ph}$ was found to be 2.5 ps. 

Fig.3(c) and Fig.3(d) show the calculated carrier and phonon temperatures and phonon occupation numbers plotted as a function of time after photoexcitation for photoexcited electron-hole densities of $2\times 10^{11}$ cm$^{-2}$ and  $7\times 10^{11}$ cm$^{-2}$. $\tau_{ph}$ was assumed to be 2.5 ps. We assumed that the phonons were initially at room temperature (300K) and the equilibrium electron density was $10^{11}$ cm$^{-2}$. In the first 500 fs, the electron temperature rapidly decreases and the phonon temperature rapidly increases. For photoexcited carrier densities in the 10$^{11}$-2$\times10^{12}$ range, the maximum phonon temperatures were found to be in the 550-1550 K range, in agreement with the earlier findings of Song et.\ al. \cite{Heinz08}. As the temperature of the carriers and the phonons approach each other, the net energy exchange between the electrons and the phonons also decreases. The hot optical phonons then become the main bottleneck for further carrier cooling. It should be noted that the subsequent cooling of the coupled carrier-phonon system is not determined by the phonon lifetime alone. The coupling of the hot carriers to the phonons retards the cooling of the phonons. Fig.3(d) shows that the phonon occupation numbers exhibit decay times that are slightly longer than the value of $\tau_{ph}$ and in the 2.9-3.5 ps range. Because the temperatures of the $\Gamma$ and K(K') point optical phonons remain close to each other, larger-energy $\Gamma$-point phonons exhibit slightly smaller occupation numbers and faster decays compared to the K(K')-point phonons. 

The results from Samples A through D show that the optical phonon cooling rates on short time scales are independent of the growth technique, the number of graphene layers, and the type of the substrate. These results imply that on the fast times scales relevant to our experiments energy transport among the graphene layers, or between the graphene layers and the substrate, is not the bottleneck for hot optical phonon relaxation. The values of the optical phonon lifetime measured in this work agree very well with the theory based on anharmonic phonon interactions developed by Bonini et.\ al. \cite{Bonini07}. Contrary to expectations \cite{Fratini08}, the substrate (SiC and quartz) optical phonons do not seem to play noticeable or distinctive role in hot carrier relaxation. This could be because the samples used in this study were not single-layer. 

The authors acknowledge support from Eric Johnson through the National Science Foundation, the DARPA Young Faculty Award, the AFOSR Contract No. FA9550-07-1-0332, monitor Donald Silversmith, and the Cornell Center for Materials Research (CCMR) program of the National Science Foundation (Cooperative Agreement No. 0520404).

%\newpage

%\newpage

%\newpage

%FIG. 1: Measured differential probe transmissions (normalized to peak value) for different graphene samples (circles). the autocorrelation of the $\sim$100 fs pump and probe pulses (dashed lines), and the best-fit theoretical curves are plotted (solid lines). The extracted values of the average optical phonon lifetime $\tau_{ph}$ are indicated for each graphene sample.

%FIG. 2: a) The electron intraband scattering time due to optical photon scattering is plotted as a function of the electron energy \cite{Rana09}. (b) The phonon Brillouin zone is shown together with the annular regions (drawn not to scale) at the $\Gamma$ and K(K') points in which the optical phonons participate in electron-phonon scattering.

%FIG. 3: (a) Calculated distribution in energy of the electron-phonon scattering events is plotted for values of $T_{c}$ equal to 1250K, 1500K, 1750K, and 2000K. Total electron density is assumed to be $7\times 10^{11}$ cm$^{-2}$. Phonon temperature is fixed at 900K. (b) Measured differential probe transmissions (normalized to the peak value) (circles) and the best-fit theoretical curves (lines) are plotted for different pump pulse energies (2.5, 6, 9.8 nJ). (c) and (d) The calculated electron and phonon temperatures and the phonon occupation numbers are plotted as a function of time after photoexcitation. The photoexcited electron-hole densities are assumed to be  $2\times 10^{11}$ cm$^{-2}$ and  $7\times 10^{11}$ cm$^{-2}$. The equilibrium electron density is assumed to be  $10^{11}$ cm$^{-2}$.


\begin{thebibliography}{99}
\bibitem{Shepard08} I. Meric, M. Y. Han, A. F. Young, B. Ozyilmaz, P. Kim, K. L. Shepard, Nature Nanotech., 3, 654 (2008).  
\bibitem{Jena09} K. Tahy, S. Koswatta, T. Fang, Q. Zhang, H. Xing, D. Jena, Proc. Dev. Res. Conf. 2009, 207 (2009).  
\bibitem{Rana08} F. Rana, IEEE Trans. Nanotechnol. 7, 91 (2008)
\bibitem{Avouris09} F. Xia, T. Mueller, R. Golizadeh-Mojarad, M. Freitag, Y. Lin, J. Tsang, V. Perebeinos, P. Avouris, Nano Lett., 9, 1039 (2009).
\bibitem{Mac09} R. Bistritzer, A.H. MacDonald, arXiv:0906.2992 (2009).
\bibitem{Bonini07} N. Bonini, M. Lazzeri, N. Marzari, F. Mauri, Phys. Rev. Lett., 99, 176802 (2007). 
\bibitem{Heinz08} D. Song, F. Wang, G. Dukovic, M. Zheng, E. D. Semke, L. E. Brus, T. F. Heinz, Phys. Rev. Lett., 100, 225503 (2008). 
\bibitem{Kampfrath05} T. Kampfrath, L. Perfetti, F. Schapper, C. Frischkorn, M. Wolf, Phys. Rev. Lett., 95, 187403 (2005). 
\bibitem{Heer06} C. Bergre, Z. Song, X. Li, X. Wu, N. Brown, C. Naud, D. Mayou, T. LI, J. Hass, A. N. Marchenkov, E. H. Conrad, P. N. First, and W. A. de Heer, Science, 312, 1191 (2006).
\bibitem{Hong09} K. S. Kim, Y. Zhao, H. Jang, S. Y. Lee, J. M. Kim, K. S. Kim, J. Ahn, P. Kim, J. Choi and B. H . Hong, Nature, 457, 706-710 (2009). 
\bibitem{Reina09} A. Reina, X. Jia, J. Ho, D. Nezich, H. Son, V. Bulovic, M. S. Dresselhaus, J. Kong, Nano Lett., 9, 30 (2009).  
\bibitem{Jahan08} J. M. Dawlaty, S. Shivaraman, M. Chandrasekhar, F. Rana, and M. G. Spencer, Appl. Phys. Lett. 92, 042116 (2008).
\bibitem{Paul08} P. A. George, J. Strait, J. Dawlaty, S. Shivaraman, Mvs Chandrashekhar,F. Rana, and M. G. Spencer, Nano Lett. 8, 4248(2008)
\bibitem{Driel09} R. W. Newson, J. Dean, B. Schmidt, and H. M. van Driel, Opt. Exp. 17(4), 2326 (2009)
\bibitem{Elsaesser09} M. Breusing, C. Ropers, T. Elsaesser, Phys. Rev. Lett., 102, 086809 (2009). 
\bibitem{Rana09} F. Rana, P. A. George, J. H. Strait, J. Dawlaty, S. Shivaraman, Mvs Chandrashekhar, and M. G. Spencer, Phys. Rev. B, 79, 115447 (2009). 
\bibitem{Piscanec07} S. Piscanec, M. Lazzeri, J. Robertson, A. C. Ferrari, F. Mauri, Phys. Rev. B, {\bf 75}, 035427 (2007).
\bibitem{Butscher07} S. Butscher, F. Milde, M. Hirtschulz, E. Malic, A. Knorr, Appl. Phys. Lett., 91, 203103 (2007). 
\bibitem{Fratini08} S. Fratini, F. Guinea, Phys. Rev. B, 77, 195415 (2008)
\end{thebibliography}
\end{document}